\documentstyle[12pt,epsf]{article}
\renewcommand{\bar}[1]{\overline{#1}}

\newcommand{\mathbf}{{\bf}}  
\newcommand{\mathrm}{{\rm}}  

\begin{document}

\begin{flushright} 
CPT-97/P.3539 \\ BIHEP-TH-97-13 \\USM-TH-70
\end{flushright}

\bigskip\bigskip
\centerline{\Large \bf The Quark Spin Distributions 
of the Nucleon}

\vspace{22pt}

\centerline{\bf Bo-Qiang Ma$^{a}$, Ivan Schmidt$^{b}$ 
and Jacques Soffer$^{c}$}

\vspace{8pt}
{\centerline {$^{a}$CCAST (World Laboratory), 
P.O.~Box 8730, Beijing 
100080, China}}

{\centerline {and Institute of High Energy Physics, 
Academia Sinica,
P.~O.~Box 918(4),} 

{\centerline {Beijing 100039, China}} 


{\centerline {$^{b}$Departamento de F\'\i sica, 
Universidad T\'ecnica Federico 
Santa Mar\'\i a,}} 

{\centerline {Casilla 110-V, 
Valpara\'\i so, Chile}}


{\centerline {$^{c}$Centre de Physique Th\'eorique, 
CNRS, Luminy Case 907,}} 

{\centerline  {F-13288 Marseille Cedex 9, France}}



\begin{abstract}
The quark helicity measured in polarized deep inelastic
scattering is different from the quark spin 
in the rest frame of the nucleon. 
We point out that the quark spin distributions 
$\Delta q_{RF}(x)$ are connected
with the quark helicity distributions $\Delta q(x)$ and the
quark transversity distributions $\delta q(x)$ 
by an approximate relation:
$\Delta q_{RF}(x) + \Delta q(x)=2 \delta q(x)$. This relation
will be useful in order to measure  
the rest frame (or quark model) spin distributions of the 
nucleon 
once the quark helicity distributions and 
quark transversity
distributions are themselves measured. We also calculate
the $x$-dependent quark transversity distributions $\delta q(x)$
and quark spin distributions $\Delta q_{RF}(x)$ in
a light-cone SU(6) quark-spectator model, and present discussions
on possible effect from the sea quark-antiquark pairs.
\end{abstract}

\vfill

\centerline{To appear in Physics~Letters B} 
\newpage

\section{Introduction}

The spin content of the proton 
has received increasing attention since the
observation of the Ellis-Jaffe sum rule violation in 
the experiments of polarized deep inelastic scattering (DIS)  
of leptons on the nucleon
at CERN, DESY and SLAC  
(For recent reviews on the data see \cite{Refs}). 
 From the naive quark model we know that the three
valence quarks provide the quantum numbers of the
proton, thus the sum of the quark spins should be equal
to the proton spin. However, it was found from the
observed value of the Ellis-Jaffe integral 
that the sum of quark helicities
is much smaller than 
1/2. This gave rise to the proton ``spin puzzle''
or ``spin crisis" since one usually 
identifies the ``quark helicity''
observed in polarized DIS with the ``quark spin''. 
However, it has been
pointed out in Ref.~\cite{Ma91b,Bro94} 
that the quark helicity ($\Delta q$) 
observed in polarized 
DIS is actually the quark spin defined in the light-cone formalism
and it is different from the quark spin ($\Delta q_{RF}$)
as defined in the quark model (or rest frame of the nucleon). 
Thus the small quark helicity sum
observed in polarized DIS is not necessarily in contradiction with
the quark model in which the proton spin is provided by the
valence quarks \cite{Ma96}.

 From another point of view, the sea quarks of the nucleon seem to 
have non-trivial non-perturbative properties \cite{Bro96} 
that may be related to a
number of empirical anomalies such as the Gottfried sum rule
violation 
\cite{NMC91}, 
the strange quark content of the nucleon
\cite{CTEQ93,CCFR}, 
the large charm quark content at high $x$
\cite{Bro81},  
as well 
as the Ellis-Jaffe sum rule violation.
There are also indications that the gluons play an important
role in the spin content of the proton \cite{Gluon}. 
Therefore the situation 
concerning the spin content 
of the proton might be more complicated
than the naive quark model picture 
in which the spin of the proton is
carried by the three valence quarks.

It would be helpful in order to clarify 
this situation if one could find
a way to measure $\Delta q_{RF}$, 
the quark spin in the rest frame of 
the nucleon (or quark model). 
It is the purpose of this paper to
point out 
an approximate relation that can be used to measure $\Delta q_{RF}$:
\begin{equation} 
\Delta q_{RF}(x)+\Delta q(x) = 2 \delta q(x),
\label{eq1}
\end{equation}
where $\Delta q(x)$ and $\delta q(x)$ are the corresponding quark 
helicity and transversity distributions, 
related to the axial quark current 
$\bar q \gamma^{\mu} \gamma^5 q$  and the tensor quark current
$\bar q \sigma^{\mu\nu} i \gamma^5 q$ \cite{h1}
respectively.
We recall that the quark helicity distributions $\Delta q(x)$ are extracted from the spin-dependent structure functions $g_1^N(x)$, defined as $g_1^N (x)=1/2\sum_{q}e_q^2\Delta q(x)$,~obtained in several polarized Deep Inelastic
Scattering experiments\cite{NSMCN}. The transversity distribution $\delta q(x)$
measures the difference of the number of quarks with transverse polarization 
parallel and antiparallel to the proton transverse polarization. It can be obtained, in principle, by measuring a Drell-Yan process in a $pp$ collision
where both protons are transversely polarized \cite{h1,Bou95,Bar97}, but
it seems rather difficult and a different method has been proposed \cite{Jaf97}. Assuming that $\Delta q(x)$ and 
$\delta q(x)$ have been measured, we can then obtain the quark spin distributions
$\Delta q_{RF}(x)$ by using Eq.~(\ref{eq1}).
We will show how
Eq.~(\ref{eq1}) can be derived 
by making use of the Melosh-Wigner rotation
connecting the ordinary quark spin and the light-cone quark spin.
We will also make numerical predictions of the $x$-dependent 
distributions $\delta q(x)$ and $\Delta q_{RF}(x)$ in
a light-cone SU(6)
quark-spectator model
and present some relevant discussions on the effect from the sea
quark-antiquark pairs.

\section{The Melosh-Wigner rotation}

It is proper to describe deep inelastic
scattering 
as the sum of incoherent scatterings
of the incident lepton on the partons 
in the infinite momentum frame or in the light-cone formalism.
We will work along with the
developements in refs.~\cite{Ma91b,Bro94,Sch97}, by taking into
account the effect due to 
the Melosh-Wigner rotation \cite{MW,MW2} which
is an important ingredient in the light-cone formalism \cite{Bro97}. 
The axial charge 
$\Delta q=\int {\mathrm d} x \Delta q(x)$  
measured in polarized deep inelastic
scattering is defined by the axial current matrix element
\begin{equation}
\Delta q=<p,\uparrow|\overline{q} \gamma^{+} \gamma_{5} q|p,\uparrow>.
\end{equation}
In the light-cone or quark-parton descriptions,
$\Delta q (x)=q^{\uparrow}(x)-q^{\downarrow}(x)$,
where $q^{\uparrow}(x)$ and $q^{\downarrow}(x)$ are the probability
of finding a quark or antiquark with longitudinal momentum
fraction $x$ and polarization parallel or antiparallel
to the proton helicity in the infinite momentum frame.
However, 
in the nucleon rest frame 
one finds \cite{Ma91b,Bro94},
\begin{equation}
\Delta q (x)
=\int [{\mathrm d}^2{\mathbf k}_{\perp}] M_q(x,{\mathbf k}_{\perp}) 
\Delta q_{RF} (x,{\mathbf k}_{\perp}),
\label{Melosh1}
\end{equation}  
with
\begin{equation}
M_q(x,{\mathbf k}_{\perp})=\frac{(k^+ +m)^2-{\mathbf k}^2_{\perp}}
{(k^+ +m)^2+{\mathbf k}^2_{\perp}},
\label{eqM1} 
\end{equation}
where $M_q(x,{\mathbf k}_{\perp})$ 
being the contribution from the relativistic effect due to
the quark transverse motions (or Melosh-Wigner rotation
effect),~ $q_{s_z=\frac{1}{2}}(x,{\mathbf k}_{\perp})$ and
$q_{s_z=-\frac{1}{2}}(x,{\mathbf k}_{\perp})$ 
being the probabilities
of finding a quark and antiquark with rest mass $m$
and transverse momentum ${\mathbf k}_{\perp}$ 
and with spin parallel and anti-parallel to the rest proton
spin, one then has, $\Delta q_{RF} (x,{\mathbf k}_{\perp})=
q_{s_z=\frac{1}{2}}(x,{\mathbf k}_{\perp})-
q_{s_z=-\frac{1}{2}}(x,{\mathbf k}_{\perp})$, 
and $k^+=x {\cal M}$, where 
${\cal M}^2=\sum_{i}(m^2_i+{\mathbf k}^2_{i \perp}) / {x_i}$.
The Melosh-Wigner rotation factor 
$M_q(x,{\mathbf k}_{\perp})$ ranges
from 0 to 1; thus $\Delta q$ measured 
in polarized deep inelastic scattering cannot be  
identified
with $\Delta q_{RF}$, the spin carried by each quark flavor 
in the proton
rest frame or the quark spin in the quark model. 

The same technique by making use 
of the Melosh-Wigner rotation effect
has been applied to the quark tensor charge  
\cite{Sch97} 
which is calculated from 
\begin{equation}
2\delta
q=<p,\uparrow|\bar{q}_{\lambda}\gamma^{\perp} \gamma^{+}
q_{-\lambda}|p,\downarrow>, 
\end{equation} 
with $\lambda=+$ and $\gamma^{\perp}=\gamma^1+i \gamma^2$,
and it is found that the quark transversity
distribution equals to 
\begin{equation}
\delta q (x)
=\int [{\mathrm d}^2{\mathbf k}_{\perp}] 
{\widetilde M}_q(x,{\mathbf k}_{\perp}) 
\Delta q_{RF} (x,{\mathbf k}_{\perp}),
\label{Melosh2}
\end{equation}  
with
\begin{equation}
{\widetilde M}_q(x,{\mathbf k}_{\perp})=\frac{(k^+ +m)^2}
{(k^+ +m)^2+{\mathbf k}^2_{\perp}}
\label{eqM2} 
\end{equation}
being the correction factor from the Melosh-Wigner rotation
\footnote{In Eq.~(\ref{eqM2}), 
$\widetilde M_q(x,{\mathbf k}_{\perp})$ 
has additional terms like
$k_{1}^2 - k_{2}^2 $
in the numerator,
where 
${\mathbf k}_{\perp}=(k_1,k_2)$
is the transverse momentum of the struck quark.
These terms vanish upon integration over the azimuth of ${\mathbf
k}_{\perp}$.}. 
 From Eqs.~(\ref{eqM1})
and (\ref{eqM2}) one easily finds the relation\cite{Sch97}
\begin{equation}
1 + M_q = 2\widetilde{M}_q.
\label{eq1b}
\end{equation}
Combining Eqs.~(\ref{Melosh1}), (\ref{Melosh2}), and (\ref{eq1b}),
one has Eq.~(\ref{eq1}).

Eq.~(\ref{eq1b}) is valid in a quite general framework of the
light-cone quark model \cite{Bro94,MW2}, and is in fact
non-perturbative. 
We point out that correction factors similar to $M_q$ and
${\widetilde M}_q$ have been also found in other papers
\cite{Bar97,Mul97} on the
quark distribution functions $\Delta q(x)$ and $\delta q(x)$. Though
the explicit expressions for the $M_q$ and ${\widetilde M}_q$ 
and the physical significances are different, Eq.~(\ref{eq1b})
also holds in these different approaches. 
Recently there has been also a proof of the above suggested relation
Eq.~(\ref{eq1}) in a QCD Lagrangian based formalism\cite{Qing98}.
Thus  
Eq.~(\ref{eq1b}), and consequently its extension to Eq.~(\ref{eq1}), 
might be
considered as a 
relation with general physical 
implications. Since $\Delta q(x)$ and $\delta q(x)$ have different
evolution behaviors, 
the relation Eq.~(\ref{eq1}) should be considered 
as valid at some model energy
scale $Q^2_0$ \cite{Bar97}, such as 
$Q^2_0 \approx 1 \to 5$ GeV$^2$ in our case.
Although it has a similar appearance, Eq.~(\ref{eq1}) 
is not a saturation
of the inequality \cite{Sof95}:
\begin{equation}
q(x) + \Delta q(x) \ge 2\big |\delta q(x)|,
\label{Sie}
\end{equation}
since $\Delta q_{RF}(x)$ is clearly not the same as $q(x)$.

\section{The light-cone SU(6) quark-spectator model}

We now discuss the $x$-dependent quark distributions 
$\Delta q_{RF}(x)$
and $\delta q(x)$ in a light-cone SU(6) quark-spectator model
\cite{Ma96},
which can be considered as a revised version of the quark-spectator
model developed in \cite{Car75}.    
The unpolarized valence quark distributions $u_v(x)$ and $d_v(x)$
are given in this model by
\begin{eqnarray} 
&&u_{v}(x)=\frac{1}{2}a_S(x)+\frac{1}{6}a_V(x);\nonumber\\
&&d_{v}(x)=\frac{1}{3}a_V(x),
\label{eq:ud}
\end{eqnarray}
where $a_D(x)$ ($D=S$ for scalar spectator or $V$ for 
axial vector
spectator) is normalized such
that $\int_0^1 {\mathrm d} x a_D(x)=3$, 
and it denotes the amplitude for quark
$q$ to be scattered while the spectator is in the diquark state $D$.
Exact SU(6) symmetry provides the relation $a_S(x)=a_V(x)$,
which implies the valence flavor symmetry $u_{v}(x)=2 d_{v}(x)$. This
gives the prediction $F^n_2(x)/F^p_2(x)\geq 2/3$ for
all $x$,  
which is ruled out by the experimental
observation $F^n_2(x)/F^p_2(x) <  1/2$ for $x \to 1$.
The mass
difference between the scalar
and vector spectators can reproduce the $u$ and $d$ valence
quark asymmetry that
accounts for the observed ratio $F_2^{n}(x)/F_2^{p}(x)$ at large $x$
\cite{Ma96}. 
This
supports the quark-spectator picture of deep inelastic scattering
in which the difference between the mass of the scalar and vector
spectators is important in order to reproduce the explicit
SU(6) symmetry breaking, while the bulk SU(6) symmetry of the
quark model still holds.

 From the above discussions 
concerning the Melosh-Wigner rotation effect,
we can write the quark helicity distributions
for the $u$ and $d$ quarks as \cite{Ma96}
\begin{eqnarray}
&&\Delta u_{v}(x)=u_{v}^{\uparrow}(x)-u_{v}^{\downarrow}(x)=
-\frac{1}{18}a_V(x)M_V(x) \nonumber \\
&&\phantom{..............................................}
+\frac{1}{2}a_S(x)M_S(x);\nonumber\\
&&\Delta d_{v}(x)=d_{v}^{\uparrow}(x)-d_{v}^{\downarrow}(x)
=-\frac{1}{9}a_V(x)M_V(x),
\label{eq:sfdud}
\end{eqnarray}
in which $M_S(x)$ and $M_V(x)$ are 
the Melosh-Wigner correction factors
for the scalar and axial vector spectator-diquark cases. 
They are obtained by averaging Eq.~(\ref{eqM1})
over ${\mathbf k}_{\perp}$ with
${\cal M}^2=(m^2_q+{\mathbf k}^2_{\perp}) /{x}
+(m^2_D+{\mathbf
k}^2_{\perp}) /{x}$, where $m_D$ is the mass of the diquark spectator,
and are unequal due to unequal spectator masses $\rightarrow$ unequal
${\mathbf k}_{\perp}$ distributions.
 From Eq.~(\ref{eq:ud}) one gets
\begin{eqnarray} 
&&a_S(x)=2u_v(x)-d_v(x);\nonumber\\
&&a_V(x)=3d_v(x).
\label{eq:qVS}
\end{eqnarray}
Combining Eqs.~(\ref{eq:sfdud}) and (\ref{eq:qVS}) we have
\begin{eqnarray} 
&&\Delta u_{v}(x)
    =[u_v(x)-\frac{1}{2}d_v(x)]M_S(x)-\frac{1}{6}d_v(x)M_V(x);
\nonumber    \\
&&\Delta d_{v}(x)=-\frac{1}{3}d_v(x)M_V(x).
\label{eq:dud}
\end{eqnarray}
Thus we arrive at simple relations 
\cite{Ma96} between the polarized
and unpolarized quark distributions for the valence $u$ and $d$
quarks. 
The relations (\ref{eq:dud})
can be considered as the results of the conventional
SU(6) quark model, and which 
explicitly take into account the Melosh-Wigner rotation effect
\cite{Ma91b,Bro94}
and the flavor asymmetry introduced by the
mass difference between the scalar and vector
spectators \cite{Ma96}.

The extension of relations Eq.~(\ref{eq:dud}) to the
quark spin distributions $\Delta q_{RF}(x)$ and transversity
$\delta q(x)$
is straightforward: we can simply replace $M_S(x)$ and $M_V(x)$
by $1$ for $\Delta q_{RF}(x)$ and by ${\widetilde M}_S(x)$ and
${\widetilde M}_V(x)$ for $\delta q(x)$,  
\begin{eqnarray} 
&&\Delta u^{RF}_{v}(x)
=u_v(x)-\frac{2}{3}d_v(x); \nonumber\\
&&\Delta d^{RF}_{v}(x)=-\frac{1}{3}d_v(x);
\label{eq:dudRF}
\end{eqnarray}
\begin{eqnarray} 
&&\delta u_{v}(x)
    =[u_v(x)-\frac{1}{2}d_v(x)]{\widetilde M}_S(x)
-\frac{1}{6}d_v(x){\widetilde M}_V(x); \nonumber \\
&&\delta d_{v}(x)=-\frac{1}{3}d_v(x){\widetilde M}_V(x).
\label{eq:dudT}
\end{eqnarray}
We notice that 
the quark spin distributions
$\Delta q_{RF}(x)$, i.e., Eq.~(\ref{eq:dudRF}), are connected
with the unpolarized quark distributions 
without any model parameter.
Thus any evidence for the
invalidity of Eq.~(\ref{eq:dudRF}), by combining together the
measured $\Delta q_v(x)$ and $\delta q_v(x)$, 
will provide a clean signature for  
new physics beyond the SU(6) quark model.

The $x$-dependent Melosh-Wigner rotation factors
$M_S(x)$ and $M_V(x)$ have been calculated \cite{Ma96} 
and an asymmetry between $M_S(x)$ 
and $M_V(x)$ was found.
The calculated polarization asymmetries
$A_1^N=2 x g_1^N(x)/F_2^N(x)$ 
including the Melosh-Wigner rotation
have been found \cite{Ma96} to be in reasonable agreement
with the experimental data, at least for $x \geq 0.1$.
A large asymmetry between $M_S(x)$ and $M_V(x)$
leads to a better fit to the data, than that 
obtained
from a small asymmetry. 
Therefore it is
reasonable to expect that the 
calculated $\delta q(x)$ and $\Delta q_{RF}(x)$ 
may lead to predictions close to the real situation.
In Fig.~(\ref{eq1}) we present 
the calculated $\Delta q(x)$, $\delta q(x)$
and $\Delta q_{RF}(x)$ for the $u$ and $d$ valence quarks.
 From Eqs.~(\ref{Melosh1}), (\ref{Melosh2}) and Fig.~(\ref{eq1}) 
we observe the inequalities,
\begin{equation}
|\Delta q_{RF}(x)| \ge |\delta q(x)| \ge |\Delta q(x)|.
\label{IE}
\end{equation}
However, the different evolution behaviors of $\delta q(x)$ and
$\Delta q(x)$ may break the inequality $|\delta q(x)| \ge |\Delta
q(x)|$ at large $Q^2$ \cite{Bar97}.
This interesting  hierarchy is specific of this model 
and is not necessarily satisfied in general.

\vspace{0.5cm}
\begin{figure}[htb]
\begin{center}
\leavevmode {\epsfysize=10cm \epsffile{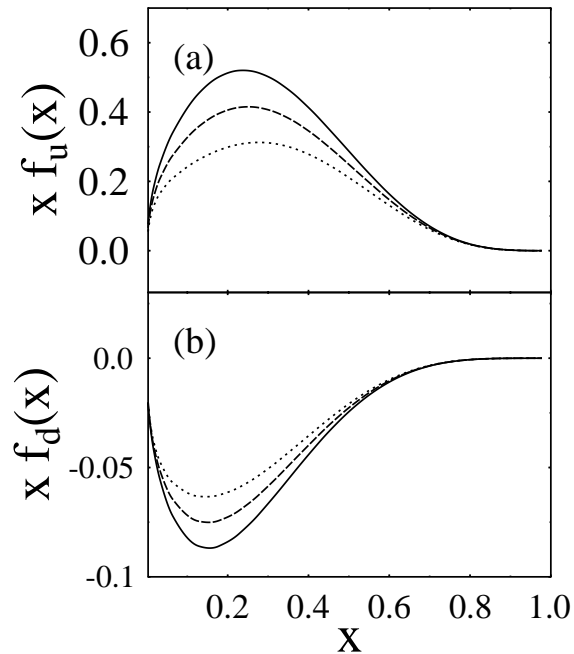}} 
\end{center}
\caption[*]{\baselineskip 13pt 
The x-dependent quark spin distributions 
$x \Delta q_{RF} (x)$ (solid curves),
transversity distributions $x \delta q(x)$ (dashed curves), and 
helicity distributions $x \Delta q(x)$ 
(dotted curves) in the light-cone
SU(6) quark-spectator model by using
Eqs.~(\ref{eq:dud}-\ref{eq:dudT}), 
with the Gl\"uck-Reya-Vogt parameterization 
\cite{GRV95} 
of unpolarized quark distributions as input:
(a) for $u$ quarks; (b) for $d$ quarks. }
\label{mssf1}
\end{figure}

As we have pointed out, one should not confuse 
Eq.~(\ref{eq1}) with the saturation of the 
inequality (\ref{Sie}), 
which is valid for each flavor,
likewise for antiquarks.  Eq.~({\ref{eq1}) only equals to the
saturated (\ref{Sie}) for the scalar spectator case, but not for
the vector spectator case due to the fact
that $q(x) \neq \Delta q_{RF}(x)$. Since 
$|\Delta q_{RF} (x)| \leq q(x)$, we may re-write from Eq.~(1) 
another inequality
\begin{equation}
q(x) \geq |2 \delta q(x) - \Delta q(x) |\ ,
\label{Sieb}
\end{equation}
which is similar to, but different from, the inequality (\ref{Sie}). 
Actually (\ref{Sie}) is a stronger constraint
than (\ref{Sieb}).
Nevertheless, we point out without detailed argument here 
that the inequality
(\ref{Sie}) is valid in the light-cone SU(6) quark model, 
even when the meson-baryon fluctuations, that will be considered in
the next section, are also taken into account.

Since the Melosh-Wigner rotation factor $M$ 
is less than 1, we expect to
find that $\sum_{q} \Delta q_{RF}$, where $\Delta q_{RF}$ is the first
moment of $\Delta q_{RF}(x)$, will be much closer to 1 than the
usual helicity sum $\Delta\Sigma = \sum_{q} \Delta q$, which
experimentally is about $0.2$, and whose departure from
the quark model value of 1 originated the ``spin crisis''.
In this context it is interesting to
notice that lattice QCD calculations gave an axial charge 
$\Delta \Sigma =0.18 \pm 0.10$ \cite{lQCD1} and
a tensor charge $\delta \Sigma =0.562 \pm 0.088$ \cite{lQCD2}. Thus
the spin carried by quarks from lattice QCD should be
$\sum_{q} \Delta q_{RF}= 0.94 \pm 0.28$
from Eq.~(\ref{eq1}), and this supports
the naive quark picture that the spin of the proton is
mostly carried by quarks.
In a quark model that does not contain antiquarks, 
$\sum_{q} \Delta q_{RF}$
will be strictly 1, but in general it will receive contributions
other than the usual valence quarks. Thus it will be of great
interest to develop a more refined quark model which can explain
or predict its actual experimental value.

\section{The sea quark-antiquark pairs}

We still need to
consider the higher Fock states 
for a better understanding of
a number of empirical anomalies related to the
nucleon sea quarks probed in deep inelastic scattering.
The Ellis-Jaffe sum rule violation is closely related to the
Gottfried sum rule violation, which implies an excess of
$d \bar d$ pairs over $u \bar u$ pairs in the
proton sea \cite{NMC91,Pi}. 
This can be explained by the  meson-baryon 
fluctuation picture of the nucleon sea \cite{Bro96,Pi}:
the lowest nonneutral $u \bar u$ fluctuation in
the proton is
$\pi^{-}(d \bar u)\Delta^{++}(uuu)$,  and its
probability is small compared 
to the less massive nonneutral $d \bar d$
fluctuation $\pi^{+}(u \bar{d})n(udd)$. 
Therefore the dominant nonneutral light-flavor $q \bar
q$ fluctuation in the proton sea is $d \bar d$ through the
meson-baryon configuration
$\pi^{+}(u \bar{d})n(udd)$. 
For the spin structure of the $q \bar q$ pairs from
the meson-baryon fluctuation model, it is observed
\cite{Bro96} that the net $d$ quark
spin of the intrinsic $q \bar q$ fluctuation is negative,
whereas the net $\bar d$ antiquark spin is zero. 

The quark helicity distributions $\Delta q(x)$ and 
transversity distributions $\delta q(x)$ 
should be measured for quarks and
antiquarks separately for applying Eq.~(\ref{eq1}). 
Thus we need techniques
that allow the measurement of $\Delta q (x)$
and $\delta q (x)$ for quarks and 
antiquarks.
The antiquark contributions to $\Delta q$ and $\delta q$ are 
predicted to be zero
in the meson-baryon fluctuation model \cite{Bro96} and in a
broken-U(3) version of the chiral quark model \cite{Che95}.
There have been explicit 
measurements of the helicity distributions
for the individual $u$ and $d$ valence 
and sea quarks by the Spin
Muon Collaboration (SMC) \cite{NSMCN}.  
The measured helicity distributions
for the $u$ and $d$ antiquarks 
are consistent with zero, 
in agreement with the above predictions \cite{Bro96,Che95}. 
The SMC data for the quark helicity distributions
$\Delta u_{v}(x)$ and $\Delta d_{v}(x)$,
which are actually $\Delta u(x)-\Delta \bar{u}(x)$
and $\Delta d(x)-\Delta \bar{d}(x)$, 
are still not 
precise enough for making detailed comparison,
but the agreement of the SMC data with 
the calculated $\Delta u_{v}(x)$ turns out to be reasonable
\cite{Ma96}.
It seems that 
there is some evidence for an additional source of negative
helicity contribution to the valence $d$ quark beyond the
conventional quark model
from the refined results by SMC \cite{NSMCN}.
This supports the 
prediction \cite{Bro96}
that the measured
$\Delta d(x) -\Delta \bar d(x)$ should receive additional negative
contribution
from the intrinsic $d$ sea quarks in comparison with the
valence-dominant 
result
presented in Fig.~(\ref{mssf1}).
 
In case of symmetric quark-antiquark sea pairs, we may
consider Eq.~(\ref{eq1}) as a relation that applies 
to valence quarks.
The tensor charge, defined as 
$\delta Q=\int_0^1 {\mathrm d} x [ \delta q(x) -\delta \bar q(x)]$,
receives only contributions from
the valence quarks since those from the sea quarks
and antiquarks cancel each other, due to the charge conjugation
properties of the tensor current 
$\bar q \sigma^{\mu\nu} i \gamma^5 q$. 
The helicity distributions for quarks and antiquarks can be
measured in semi-inclusive deep inelastic processes separately 
\cite{NSMCN}, thus we can measure
the valence quark helicity distributions 
defined by $\Delta q_v(x)=\Delta q(x)-\Delta \bar q(x)$
from experiment. 
We also notice that
there is no clear way to strictly 
distinguish between the valence quarks
and sea quarks for the $u$ and $d$ flavors, since one can
have a symmetric quark-antiquark sea pairs by defining
the valence quark 
$q_v=q-\bar q$ due to the excess of net  
$u$ and $d$ quarks in the nucleon. 
Eq.~(\ref{eq1}) is also valid for 
the above defined
valence quarks
(which should be actually $q-\bar q$)
in case of non-zero spin contribution from antiquarks.   

One 
interesting feature of the meson-baryon fluctuations
is the strange
quark-antiquark asymmetry from the 
virtual $K^+ \Lambda$ pair of the proton \cite{Bro96}. 
The intrinsic strangeness fluctuations in the
proton wavefunction are mainly 
due to the intermediate $K^+ \Lambda$
configuration since this state 
has the lowest off-shell light-cone
energy and invariant mass. 
The
intrinsic strange quark normalized to the probability
$P_{K^+\Lambda}$ of the $K^+\Lambda$ configuration yields a
fractional contribution $\Delta S_{s}=2
S_z(\Lambda)=-\frac{1}{3}P_{K^+\Lambda} $ to the proton spin,
whereas the intrinsic antistrange 
quark gives a zero contribution:
$\Delta S_{\bar s}=0$ \cite{Bro96}. 
In case of symmetric strange
quark-antiquark pairs, 
one shall predict a zero strange tensor charge.
However, a non-zero
strange tensor charge will arise from the strange
quark-antiquark spin asymmetry 
due to the meson-baryon fluctuations
and we predict a strange tensor charge
$\delta s \approx -0.02 \to -0.03$
(similar to 
$\Delta s$ \cite{Bro96}) 
corresponding to the
probability  $P_{K^+\Lambda}=5 \to 10 \%$. 

\section{Discussion and Summary}

In this paper we have proposed an approximate relation that
can be used to measure the quark spin distribution
$\Delta q_{RF}(x)$, as implied in the quark model or in the
rest frame of the nucleon.
It will be very meaningful if a clear definition of this
spin distribution or any
other way for measuring this quantity can be found. 
It has been noticed recently \cite{MS2} 
that the quark spin distribution
defined in this paper is actually equivalent to
$\Delta q(x)+2 L_q(x)$, where $\Delta q(x)$ is the quark helicity
distribution and $L_q(x)$ is the quark orbital angular
momentum obtained by calculating the matrix element of 
the operator ${\mathbf L}_q
=-i \gamma^+ {\mathbf k} \times \nabla
_{\mathbf k}$. 
Thus  $\Delta q_{RF}(x)$ is a quantity that can be calculated
in an exact theoretical framework, such as lattice QCD, 
and might be measurable in the future.  This means that
Eq.~(\ref{eq1}) might be a practical relation
that can be tested by other means. 

In summary,
we showed in this paper that the quark spin distributions
$\Delta q_{RF}(x)$, in the rest frame of the nucleon, 
are connected
with the quark helicity distributions $\Delta q(x)$ and
the quark transversity distributions $\delta q(x)$ by  
an approximate but simple
relation: $\Delta q_{RF}(x) + \Delta q(x)=2 \delta q(x)$. This 
relation
will be useful to measure  
the quark spin distributions of the nucleon  
once the quark helicity distributions and 
quark transversity
distributions are measured. 
It will be also very useful in order to 
check
various models and will provide more information concerning the
spin structure of the nucleon. 

\bigskip
{\bf Acknowledgments: } 
We would like to thank V.~Barone, S.J.~Brodsky, 
R.~Jakob, K.-F.~Liu, and P.J.~Mulders 
for helpful discussions. This work is 
partially supported by National Natural 
Science Foundation of China 
under Grant No.~19605006, Fondecyt (Chile) under grant 1960536, 
by the cooperation programme 
ECOS-CONICYT between France and Chile
under contract No. C94E04, by a C\'atedra Presidencial (Chile),
and by Fundaci\'on Andes (Chile).

\end{document}